%% file: kosa2020b.tex
\documentclass[fleqn,usenatbib]{mnras}

\usepackage{newtxtext,newtxmath}
\usepackage[T1]{fontenc}

\DeclareRobustCommand{\VAN}[3]{#2}
\let\VANthebibliography\thebibliography
\def\thebibliography{\DeclareRobustCommand{\VAN}[3]{##3}\VANthebibliography}

\usepackage{graphicx}
\usepackage{amsmath}
 
\usepackage{amssymb}

\title[The 40.5-min period eclipsing binary SDSS J082239.54$+$304857.19]{Multi-band light curve analysis of the 40.5-minute period eclipsing double-degenerate binary SDSS J082239.54$+$304857.19}

\author[Kosakowski, Kilic, \& Brown]{Alekzander Kosakowski$^{1}$,
Mukremin Kilic$^{1}$,
Warren Brown$^{2}$
\\
$^{1}$Homer L. Dodge Department of Physics and Astronomy, University of Oklahoma, 440 W. Brooks St., Norman, OK, 73019 USA\\
$^{2}$Smithsonian Astrophysical Observatory, 60 Garden St, Cambridge, MA 02138 USA\\
}

\date{Accepted XXX. Received YYY; in original form ZZZ}

\pubyear{2020}

\begin{document}
\label{firstpage}
\pagerange{\pageref{firstpage}--\pageref{lastpage}}
\maketitle

\begin{abstract}

We present the Apache Point Observatory BG40 broadband and simultaneous Gemini $r$-band and $i$-band high-speed follow-up photometry observations and analysis of the 40.5 minute period eclipsing detached double-degenerate binary SDSS J082239.54$+$304857.19. Our APO data spans over 318 days and includes 13 primary eclipses, from which we precisely measure the system's orbital period and improve the time of mid-eclipse measurement. We fit the light curves for each filter individually and show that this system contains a low-mass DA white dwarf with radius $R_A=0.031\pm0.006~{\rm R_\odot}$ and a $R_B=0.013\pm0.005~{\rm R_\odot}$ companion at an inclination of $i=87.7\pm0.2^\circ$. We use the best-fitting eclipsing light curve model to estimate the temperature of the secondary star as $T_{\rm eff}=5200\pm100~{\rm K}$. Finally, while we do not record significant offsets to the expected time of mid-eclipse caused by the emission of gravitational waves with our 1-year baseline, we show that a $3\sigma$ significant measurement of the orbital decay due to gravitational waves will be possible in 2023, at which point the eclipse will occur about $8$ seconds earlier than expected.
\end{abstract}


\section{Introduction}

Eclipsing binary systems provide rare opportunities to directly measure the physical parameters of both of the stars in the system. If the primary and secondary eclipses are both clearly visible, it is possible to test the theoretical mass-radius relationship \citep{parsons2017}. Furthermore, with precisely-measured mid-eclipse times, it is also possible to measure the effects of orbital decay due to the loss of orbital angular momentum from gravitational wave emission and torques caused by tidal interaction \citep{piro2011, benacquista2011, fuller2013}.

Even in eclipsing systems where the secondary eclipse is obscured by a significantly brighter primary star, it is still possible to place constraints on the properties of the hidden secondary star by using the information contained within the primary eclipse through light curve fitting. Additional information on these invisible companions can be obtained from radial velocity measurements of the primary star, which provide information on the system's orbital period and mass ratio. Comparing results from light curve fitting with evolutionary models and stellar atmosphere models allows for an independent way to confirm the temperature and radii of both of the stars in the binary system.

As of this work, there are only 14 known eclipsing double-degenerate systems. With periods ranging from 7 to 354 minutes, these systems are: NLTT 11748 \citep{steinfadt2010}, CSS 41177 \citep{drake2010, parsons2011}, GALEX J171708.5$+$675712 \citep{vennes2011}, J0651$+$2844 \citep{brown2011}, J0751$-$0141 \citep{kilic2014}, J1152$+$0248 \citep{hallakoun2016}, J0822$+$3048 \citep{brown2017}, J1539$+$5027 \citep{burdge2019}, ZTF J1901$+$5309 \citep{coughlin2020}, ZTF J0538$+$1953 \citep{burdge2020a}, ZTF J2029+1534 \citep{burdge2020a}, ZTF J0722$-$1839 \citep{burdge2020a}, ZTF J1749$+$0924 \citep{burdge2020a}, and ZTF J2243+5242 \citep{burdge2020b}. Here we report on follow-up observation and analysis of the relatively faint ($g_0=20.198\pm0.023~{\rm mag}$), 40.5-minute period double-degenerate eclipsing binary system SDSS J082239.54$+$304857.19 (hereafter: J0822$+$3048) using the APO 3.5-meter and Gemini North 8.1-meter telescopes.

Originally discovered by \citet{brown2017} as a part of an ongoing search for extremely low mass ($M<0.3~{\rm M}_\odot$) white dwarfs \citep{brown2020,kosakowski2020}, J0822$+$3048 is the seventh eclipsing double white dwarf binary discovered. The authors used the MMT 6.5-meter telescope with the blue-channel spectrograph to obtain radial velocity measurements of the J0822$+$3048 system and showed that it contains a $M_A=0.304\pm0.014~{\rm M_\odot}$ DA white dwarf and a degenerate companion with mass $M_B=0.524\pm0.050~{\rm M_\odot}$ on a 40.5 minute orbit. They followed-up their spectroscopic observations with 68 minutes of broadband photometry with a blue filter made of Schott BG40 filter glass (BG40 filter, $340-600~{\rm nm}$) using the Apache Point Observatory (APO) 3.5-meter telescope frame-transfer camera, Agile, and found two short ($\sim60$-second), 0.2 mag deep eclipses in the light curve with a separation consistent with the orbital period obtained through their radial velocity fits. Based on these two eclipses, the authors placed weak constraints on the absolute radii of the stars in the system.

We expand upon the discovery data with an additional 492 minutes of APO BG40 broadband filter data spread across two additional observing sessions for a total BG40 filter baseline of over 318 days, as well as 209 minutes of simultaneous $r$-band and $i$-band filter data from the 8.1-meter Gemini North telescope using the high-speed camera, 'Alopeke. We use these data to further constrain the component radii, orbital inclination, and mid-eclipse timing of the system.

This paper is organized as follows. In section 2, we present the observations and discuss the data reduction steps used to create our final light curves. In section 3 we discuss our data analysis methods, and in section 4 we discuss our results on the binary system parameters and conclude.

\section{Observations and Data Reduction}

\subsection{Apache Point Observatory}
J0822$+$3048 was originally observed on UT 2017 March 02 using the 3.5-meter telescope at the Apache Point Observatory (APO) with the BG40 broadband filter on the Agile frame-transfer camera \citep{mukadam2011} exposing for 68 minutes with 30-second back-to-back exposures. This discovery dataset captured two primary eclipses, each containing only two data points.

We obtained follow-up data on UT 2017 November 16 and UT 2018 January 14 using an identical setup to the discovery data with 30-second exposures. Our first night of follow-up observations spanned 322 minutes. We excluded the final 81 minutes of data due to cloud coverage significantly affecting the light curve. The remaining 241 minutes includes six primary eclipses. Our second night of follow-up observations spanned 251 minutes and covers six primary eclipses. One of these eclipses is lost due to instrument problems. Figure \ref{fig:raw_model_lightcurves} shows our calibrated light curves for our BG40 datasets. Our best-fitting model from our Monte Carlo light curve fits to the APO BG40 data (discussed below) is overplotted as a solid red line.

\subsection{Gemini North}
We supplemented the APO BG40 broadband filter data with simultaneous $r$- and $i$-band observations using the dual-channel frame-transfer camera, 'Alopeke \citep{scott2018} on the 8.1-meter Gemini North telescope. The observations were taken in eight, nearly back-to-back, observing blocks each containing 100 back-to-back 15-second exposures on UT 2019 March 12 as a part of the program GN-2019A-Q-119. These observations spanned 209 minutes and included five primary eclipses. Cloud coverage affected the quality of the data about 2 hours into the observations. Unfortunately, due to an issue with the GPS timing synchronization between the telescope and the 'Alopeke instrument at the time of observation, the Gemini data is systematically shifted by about $-21$ seconds. We note that the relative frame timing is unaffected by this systematic shift. Figure \ref{fig:gemini_lightcurves} shows our calibrated Gemini $r$-band (top) and $i$-band (bottom) light curves. Our best-fitting model from our Monte Carlo light curve fits to the Gemini $r$-band and $i$-band data (discussed below)  are overplotted as solid red lines.

\subsection{Data Reduction}
We used the IRAF package CCDRED to perform image reduction using a set of bias images, dark images, and twilight flats, each taken on the same nights as our observations. We performed relative aperture photometry using the IRAF package DAOPHOT using a circular source aperture with radius based on the FWHM of each image and a background annulus surrounding each source aperture. For our APO Agile data, we used two nearby, relatively bright, nonvariable field stars to calibrate the resulting light curve. For our Gemini 'Alopeke data, because the 'Alopeke instrument has a much smaller field of view than Agile, we only had three nearby, non-variable, field stars of similar brightness available to calibrate our target light curve. We detrended each light curve by fitting and subtracting a third-order polynomial. Finally, we converted our APO data timing system from beginning-of-exposure TAI to middle-of-exposure Barycentric Dynamical Time \citep[BJD\_TDB,][]{eastman2010} and our Gemini data timing from end-of-exposure TAI to middle-of-exposure BJD\_TDB.

\input{raw_model_lightcurves.tex}
\input{gemini_lightcurves.tex}

\section{Initial Period Determination}

The orbital period of the J0822$+$3048 system was originally determined using radial velocity measurements based on the Balmer lines in the optical spectrum and roughly confirmed through light curve fitting of the discovery light curve containing two adjacent primary eclipses. We combined our new APO BG40 broadband data with the discovery dataset to create a master light curve spanning just over 318 days and containing 13 primary eclipses. We use this master light curve to perform light curve fitting and to determine the orbital period of the J0822$+$3048 system.

Since the periods obtained through radial velocity measurements and the eclipsing light curve of the discovery dataset are only roughly consistent, to obtain an appropriate initial period estimate for light curve fitting, we used the AstroPy implementation of the Lomb-Scargle periodogram to create a power spectrum from all of the combined APO BG40 data using simple models with varying numbers of sine-terms. We limited our period range to search between 40 and 41 minutes with a step size of about 1 ms. Our Lomb Scargle models each returned an identical best-fit frequency at 35.55448746 cycles d$^{-1}$. While we do not estimate uncertainties on this initial measurement, this frequency is only 0.0002 cycles d$^{-1}$ ($\approx0.01~{\rm s}$) greater than the original frequency obtained using radial velocity observations of 35.55429140 cycles d$^{-1}$. We use this period as our initial value when performing light curve fitting discussed in the next section.

\section{System Parameters - Light Curve Fitting}

\subsection{APO BG40 Broadband fits}
We modeled the system parameters using JKTEBOP \citep{southworth2013}, which uses Levenberg-Marquardt minimization to obtain best-fit parameter values. For the BG40 dataset, we fit for the sum and ratio of the fractional system component radii ($r=R/a$), inclination angle, stellar light ratio, and orbital period. We chose to fix the mass ratio and initialized these parameters based on values taken from the discovery paper, with exceptions for the orbital period, which we initialized based on our previous Lomb-Scargle estimate.

We used a 4-parameter limb darkening law with coefficients for a $T_{\rm eff}=14,000~{\rm K}$, $\log{g}=7.14$ He core white dwarf primary star and $T_{\rm eff}=5,000~{\rm K}$, $\log{g}=8.00$ C/O core white dwarf secondary star. Due to technical limitations in the JKTEBOP software restricting limb darkening values to be greater than $-1.0$, we used the limb-darkening coefficients of \citet{gianninas2013} for the LSST $u$-, $g$-, $r$-, and $i$-band filters and converted these to the BG40 broadband filter system using equation 3 of \citet{hallakoun2016}. The \citet{gianninas2013} intensity functions are in good agreement with the updated \citet{claret2020} intensity functions so we expect this substitution to have minimal effect on our results. Similarly, we used fixed gravity darkening coefficients from \citet{claret2020} for the $u$-, $g$-, $r$-, and $i$-band filters and once again converted these to the BG40 system using equation 3 in \citet{hallakoun2016}. The best-fitting models for our APO BG40 and Gemini $r$-band and $i$-band fits are overplotted onto the calibrated light curve data and shown in Figures \ref{fig:raw_model_lightcurves} and \ref{fig:gemini_lightcurves}, respectively.
\input{bg40_corner_fig.tex}

We used JKTEBOP's Monte Carlo analysis to create parameter distributions and estimate uncertainties for each of our fitted parameters. This is done by creating a simulated light curve based on Gaussian perturbations to the best-fit model light curve and performing Levenberg-Marquardt minimization to the simulated light curve. Details for this Monte Carlo analysis method can be found in \citet{southworth2004} and \citet{southworth2005}. We performed 15,000 of these Monte Carlo fits to the combined BG40 master light curve and filtered out results that converged to unphysical values, such as inclination angles $i < 80^\circ$ that would not show eclipses in this system. After filtering, we were left with over 13,600 successful fits from which we created the resulting parameter distributions. Figure \ref{fig:bg40_corner} shows the final parameter distributions for our APO BG40 light curve fits. The diagonal shows the 1-D histograms with a 1-D Gaussian Kernel Density Estimate (KDE) overplotted as a blue-shaded distribution. We marked the locations of the median fit and the $1\sigma$ range of the data if the distribution is single-peaked. Because our 30-second exposures poorly sample the short primary eclipses, and because our light curves do not show a clear secondary eclipses, the secondary star's radius and the system's inclination are not well-constrained to a single best value and are strongly anti-correlated. For these double-peaked distributions, we fit a Gaussian to each peak separately and report the resulting central value and width of each Gaussian as the `best' fits. We overplot these best fits and their $1\sigma$ range as red and blue vertical lines on top of their respective peaks. Best-fit values for each parameter are reported above each histogram. The off-diagonal plots show 2-D distributions of each Monte Carlo fit with individual results marked as black points and 2-D Gaussian KDE overplotted as colored contours.

\subsection{Gemini $r$-band \& $i$-band fits}
For our Gemini $r$-band and $i$-band fits, we performed 15,000 Monte Carlo simulations fitting for sum and ratio of the fractional system component radii, inclination, and light ratio. We initialized the parameters based on the best-fitting parameters from the APO BG40 data. We chose to fix the period at the best-fit result from the BG40 data fit due to the much longer baseline of the APO data. Our Gemini $r$-band and $i$-band parameter distributions can be seen in Figures \ref{fig:gemini_r_corner} and \ref{fig:gemini_i_corner} and follow the same organization as Figure \ref{fig:bg40_corner}.

While all peak values agree within their respective $1\sigma$ ranges across each filter, we note that the large temperature difference between the primary and secondary stars resulted in a $3\sigma$ detection of the cooler secondary star in the system's light ratio for the redder Gemini $r$- and $i$-band filters. This increased significance allowed the Gemini fits to strongly favor one peak over another, essentially breaking the degeneracy between the system's inclination and the secondary star's fractional radius. Best-fit parameters and their uncertainties for all filters are presented in Table \ref{tab:all_jktebop_table}, along with their variance-weighted mean values. We calculated absolute radius values based on our light curve fitting using the orbital separation $a=0.364\pm0.008~{\rm R_\odot}$ from the discovery publication. Figure \ref{fig:phase_model_lightcurves} shows the resulting phase-folded light curves using the period from our APO BG40 dataset fits. We overplot the best-fit models created from the best-fit parameters in Table \ref{tab:all_jktebop_table} as a solid red line and zoom in to the regions surrounding the primary and secondary eclipses.

\input{all_jktebop_table3.tex}
\input{gemini_r_corner_fig.tex}
\input{gemini_i_corner_fig.tex}
\input{phased_model_lightcurves.tex}

\section{Estimating Effective Temperature and Radius of the Secondary Star}
\subsection{Temperature Estimate}

We used the best-fitting parameters from our light curve fitting to estimate the effective temperature of the secondary star. We first calculated the system's absolute magnitude using the extinction-corrected SDSS apparent magnitudes and the distance from the discovery data obtained through spectroscopic models. We then interpolated over the C/O core DA white dwarf cooling models of \citet{tremblay2011}\footnote{http://www.astro.umontreal.ca/~bergeron/CoolingModels} to a mass of $M_B=0.524~{\rm M_\odot}$. Our interpolation resulted in effective temperatures $T_{{\rm eff},r}=5210\pm150~{\rm K}$ and $T_{{\rm eff},i}=5180\pm120~{\rm K}$ for the $r$-band and $i$-band, respectively. We take the variance-weighted mean of these results and accept $T_{\rm eff}=5200\pm100~{\rm K}$ as the secondary white dwarf's effective temperature.

\subsection{Radius Estimates}

Between our light curve fitting results across three filters, the component radii for the J0822$+$3048 system are fairly-well constrained to a single solution. Here we compare our results to evolutionary model predictions for an $M=0.304\pm0.014~{\rm M_\odot}$ He core primary white dwarf and $M=0.524\pm0.050~{\rm M_\odot}$ C/O core secondary white dwarf.

For the primary white dwarf, we interpolated over the He core white dwarf evolutionary tracks of \citet{istrate2016}, including elemental diffusion and stellar rotation, and obtain a primary radius of $R_A=0.025\pm0.001~{\rm R_\odot}$. This value roughly agrees within $1\sigma$ of our estimate from the light curve fitting of $R_A=0.031\pm0.006~{\rm R_\odot}$. 

For the secondary star, we interpolated over the evolutionary models for C/O core composition, thick hydrogen layer ($q_H=10^{-4}$), white dwarfs \citep{fontaine2001}. This interpolation resulted in a radius estimate of $R_B=0.014\pm0.001~{\rm R_\odot}$. This is in excellent agreement with our estimate from light curve fitting of $R_B=0.013\pm0.005~{\rm R_\odot}$. Our light curve fitting results agree well with the mass-radius relation for white dwarfs and confirms that our Gemini $r$- and $i$-band fit results have identified the correct peak where our APO BG40 fit failed. In addition, because the secondary star's radius and the system's inclination were strongly anti-correlated, we are now also able to select the correct inclination peak at $i=87.9\pm0.4^\circ$ in our APO BG40 distribution.

\section{Eclipsing Timing Estimate and Orbital Decay}

The orbit of compact double-degenerate systems decays due to the loss of angular momentum \citep{landau1958}. While gravitational waves are generally the dominant source of angular momentum loss in these compact systems, torques caused by strong tidal interaction between the stars in compact systems may also contribute significantly to the total angular momentum loss \citep{piro2011, benacquista2011, fuller2013}. Eclipse timing measurements taken over long baselines have been used as a method to directly measure the effects of orbital decay in these systems.

In the case of the 12-minute period eclipsing double-degenerate binary J0651$+$2844 \citep{brown2011}, \citet{hermes2012} measured the system's mid-eclipse timing over a baseline of 13 months and show that the period of the system is decaying at a rate of $\dot{P}=(-9.8\pm2.8)\times10^{-12}$ s s$^{-1}$. They showed that, while the system shows evidence for tidal interaction in its ellipsoidal variations, a longer baseline is required to measure the orbital decay contribution from the tidal interaction in the system.

Similarly, \citet{burdge2019} have used new and archival data to perform mid-eclipse timing measurements of the 7-minute period eclipsing double-degenerate binary J1539$+$5027. They precisely measured the system's orbital decay with a 10 year baseline and showed that the orbital decay is consistent with constant change in period $\dot{P}=(-2.373\pm0.005)\times10^{-11}$ s s$^{-1}$. Additionally, \citet{burdge2019b} have identified a 20-minute non-eclipsing double-degenerate binary system showing strong ellipsoidal variation caused by tidal distortions. They used these ellipsoidal variations to measure the orbital decay of the system caused by gravitational wave emission and estimated the contribution to the decay from tidal effects. Finally, \citet{burdge2020b} have identified an 8.8-minute period eclipsing double-degenerate binary system using ZTF archival data and show that the system is undergoing rapid orbital decay. They estimate that tidal effects could contribute as much as $7.5$ percent to the orbital decay of the system.

Here we measure the time of mid-eclipse for each of our APO epochs to prepare for future orbital decay studies of the J0822$+$3048 system. Because of the systematic offset in the timing of our Gemini data, we estimated the mid-eclipse time only for the three epochs of APO BG40 data. For each epoch of data, we performed 50,000 Monte Carlo fits using JKTEBOP to fit the light curves for only the mid-eclipse time, using the best-fit parameters in Table \ref{tab:all_jktebop_table} as initial parameters. 
\input{eclipse_table.tex}
\input{apo_ephem_fig.tex}

We used the time of ingress and egress from the best-fit model light curve to estimate the eclipse duration as $T_{eclipse}\approx90$ seconds, with minimum light lasting $\approx20$ seconds. We therefore exclude results with mid-eclipse timing greater than 45 seconds from the median fit value, as those results place the middle of the eclipse outside of the observed range of the eclipse itself. Figure \ref{fig:apo_ephem} shows the resulting distribution for each epoch of data. We fit a Gaussian to the central peak of each distribution and report the central value and width as the best-fit and $1\sigma$ uncertainty. These values are reported in Table \ref{tab:eclipse}. We calculated the offset of each observed mid-eclipse time from its expected value by measuring the eclipse timing offset from a linear projection based on the first epoch's time of eclipse and the orbital period of the system. We note that in our second and third APO BG40 data sets, the measured mid-eclipse timings are $+3.0$ and $+2.9$ seconds off of the expected time assuming no orbital decay, but each agree within the relatively large $\pm4.3$ seconds $1\sigma$ range on the discovery dataset's mid-eclipse timing. 

Despite not recording a significant offset in the measured mid-eclipse timing, we revisited the decay of mid-eclipse timing due to gravitational waves using the two new epochs of APO data discussed in this work. Figure \ref{fig:grav_offset} shows an $(O-C)$ diagram with the best-fit mid-eclipse timing measurements to our two new epochs of APO BG40 data (black data points with error bars) plotted as an offset from the expected mid-eclipse timing assuming no orbital decay (black dashed line) with a period of 0.0281258394 d. We exclude the discovery data set due to its relatively large uncertainties. We plot the projected offsets in mid-eclipse timing out to the expected launch date of the LISA mission in 2034, based on angular momentum loss solely due to the emission of gravitational waves \citep{piro2011} as a red dashed line with shaded $1\sigma$ region dominated by the uncertainties in the masses of each star in the system. We used these projected values and the mean uncertainty in our measured values with 30-second exposure times and calculate that a $3\sigma$ significant mid-eclipse timing offset measurement will be possible in the year 2023, at which point the J0822$+$3048 system will eclipse $8.4\pm0.7$ seconds earlier than expected as measured from our second APO epoch. 

\input{grav_offset2.tex}

\section{Conclusions}

We have expanded upon the discovery APO BG40 light curve of J0822$+$3048 with an additional 492 minutes of APO BG40 data and 209 minutes of simultaneous Gemini $r$-band and $i$-band data. We analyzed these light curves and improved the estimates for the absolute radii of both stars in the system using a combination of light curve fitting and white dwarf evolutionary models. Our fits resulted in variance-weighted mean component radii values of $R_A=0.031\pm0.006~{\rm R_\odot}$, $R_B=0.013\pm0.005~{\rm R_\odot}$, and system inclination $i=87.7\pm0.2^\circ$. In addition, we use the results from light curve fitting together with white dwarf cooling models to estimate the secondary white dwarf's effective temperature at $T_{\rm eff}=5200\pm100~{\rm K}$.

Finally, we have reported an accurate and precise orbital period for this system and measured mid-eclipse times for each epoch of APO data for use with future eclipsing timing variability studies. We show that, with 30-second exposures, a $3\sigma$ significant mid-eclipse offset measurement will be possible during the year 2023, at which point the mid-eclipse time will be offset by $-8.4\pm0.7$ seconds due to the loss of angular moment from the emission of gravitational waves. 
With the expected launch of the LISA mission in 2034, we predict that J0822$+$3048 will show an $83.7\pm7.3~{\rm s}$ offset in mid-eclipse timing by the time LISA launches. While J0822$+$3048 falls just below the LISA 4-year sensitivity curve with a signal-to-noise ratio of $S/N\approx3.6$, with its precise period and sky position known, a gravitational wave detection may be possible.

While we have placed constraints on the parameters of the stars in the system, there is still room for improvement. Higher-quality data may provide the first direct detection of the secondary eclipse, allowing for absolute measurements on the secondary's radius and temperature. Additionally, reduced exposure times will allow for increased significance in future orbital decay measurements and are therefore also desired.

\section*{Acknowledgements}
This work was supported in part by the NSF under grant AST-1906379. This research is based in part on observations obtained with the Apache Point Observatory 3.5-meter telescope, which is owned and operated by the Astrophysical Research Consortium

Some of the Observations in the paper made use of the High-Resolution Imaging instrument(s) ‘Alopeke. ‘Alopeke was funded by the NASA Exoplanet Exploration Program and built at the NASA Ames Research Center by Steve B. Howell, Nic Scott, Elliott P. Horch, and Emmett Quigley. ‘Alopeke was mounted on the Gemini North telescope of the international Gemini Observatory, a program of NSF’s NOIRLab, which is managed by the Association of Universities for Research in Astronomy (AURA) under a cooperative agreement with the National Science Foundation. on behalf of the Gemini partnership: the National Science Foundation (United States), National Research Council (Canada), Agencia Nacional de Investigación y Desarrollo (Chile), Ministerio de Ciencia, Tecnología e Innovación (Argentina), Ministério da Ciência, Tecnologia, Inovações e Comunicações (Brazil), and Korea Astronomy and Space Science Institute (Republic of Korea). 

This research made use of Astropy,\footnote{http://www.astropy.org} a community-developed core Python package for Astronomy \citep{astropy2013, astropy2018}. 

\section*{Data availability}
The data underlying this article will be shared on reasonable request to the corresponding author.

\bsp
\label{lastpage}
\end{document}

%% file: raw_model_lightcurves.tex
\begin{figure*}
\centering
\includegraphics[scale=0.3]{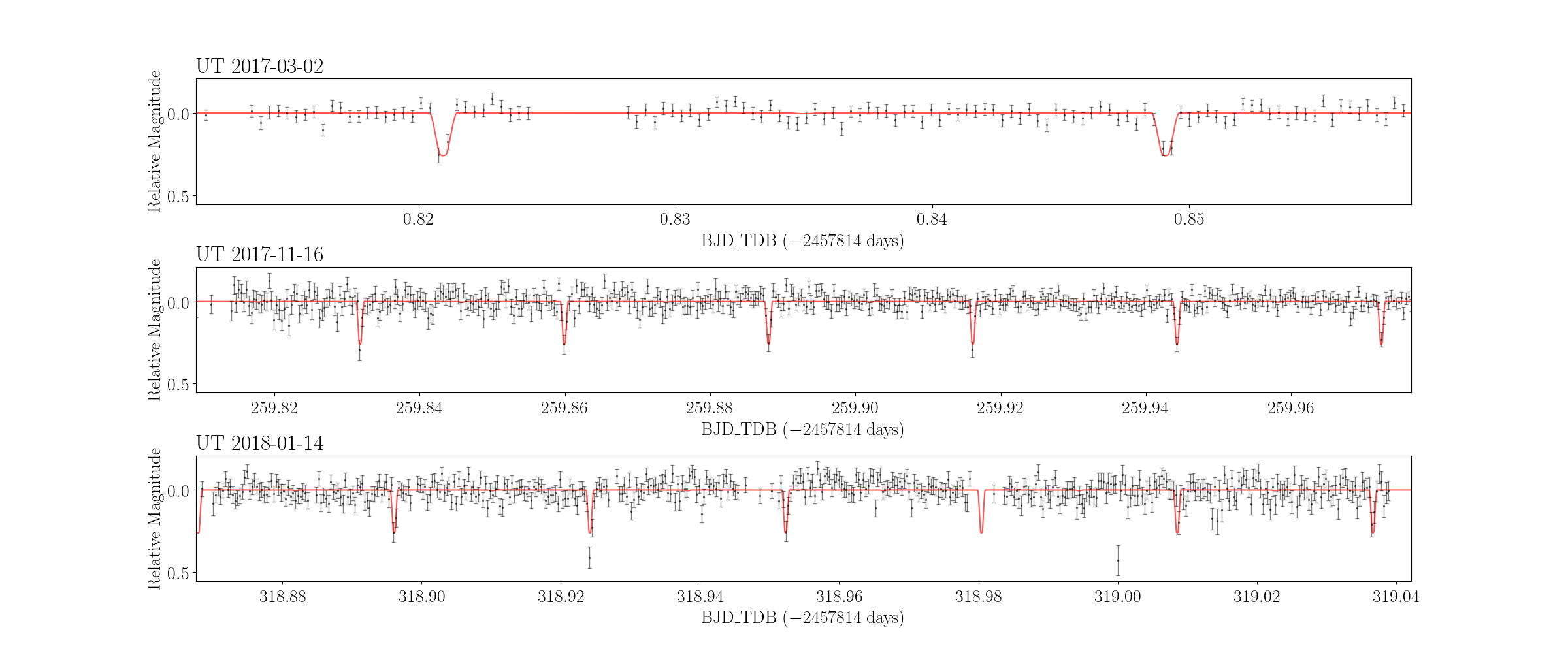}
\caption{Calibrated light curves for APO BG40 broadband from UT 2017 March 02 (top), UT 2017 November 16 (middle), and UT 2018 January 14 (bottom).
The best-fit model based on light curve fitting to the combined APO data with JKTEBOP discussed in the text is overplotted in red.}
\label{fig:raw_model_lightcurves}
\end{figure*}

%% file: gemini_lightcurves.tex
\begin{figure*}
\centering
\includegraphics[scale=0.3]{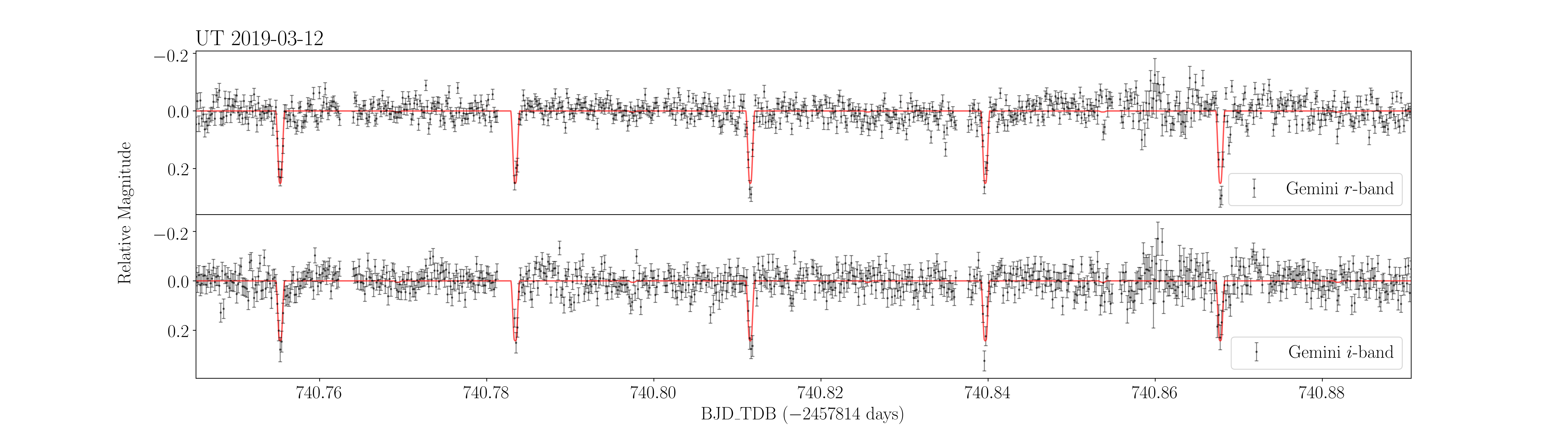}
\caption{Calibrated light curves for Gemini $r$-band (top) and $i$-band (bottom) filters obtained simultaneously on UT 2019 March 12.
The best-fit model based on light curve fitting with JKTEBOP discussed in the text is overplotted in red.
The timing shown is as recorded by the 'Alopeke instrument and is systematically offset by about 21 seconds due to an instrument problem discussed in the text.}
\label{fig:gemini_lightcurves}
\end{figure*}

%% file: bg40_corner_fig.tex
\begin{figure*}
\centering
\includegraphics[scale=0.25]{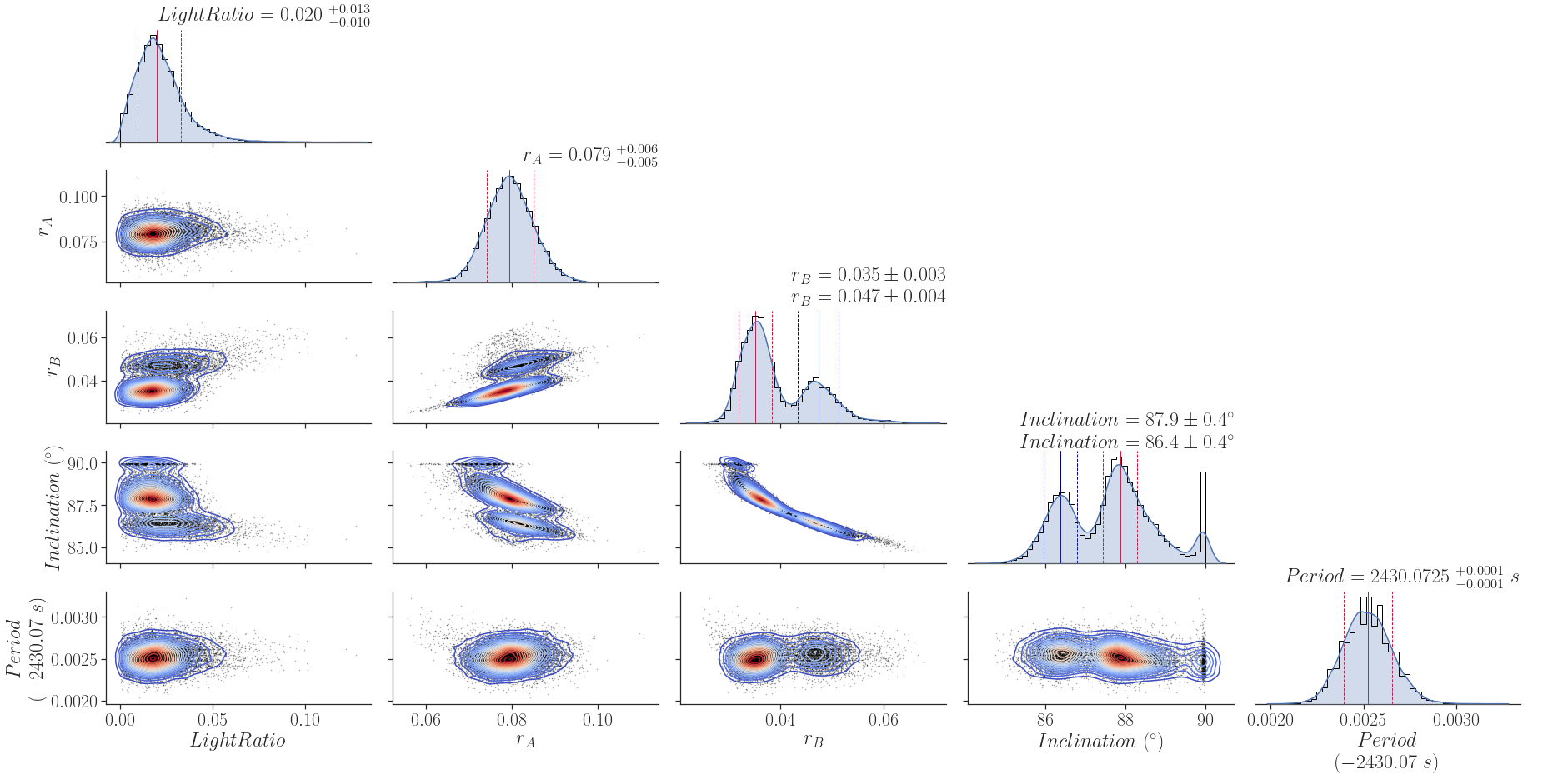}
\caption{Parameter distributions from Monte Carlo fits to the combined APO BG40 broadband light curve.
The diagonal contains the 1-D parameter distributions split into 40 bins (black histogram) with a 1-D Gaussian KDE overplotted as a blue shaded distribution.
The off-diagonal plots contain the 2-D parameter distributions with 2-D Gaussian KDE contours overplotted.
Primary/Secondary median fits and $1\sigma$ ranges are marked with vertical red/blue lines for single/double-peaked distributions.
Due to the poorly constrained light ratio and lack of visible secondary eclipses in the light curve, the secondary star's radius and the system's inclination are not well-constrained to a single peak and are strongly anti-correlated.}
\label{fig:bg40_corner}
\end{figure*}

%% file: all_jktebop_table3.tex
\begin{table*}
	\centering
  \renewcommand{\arraystretch}{1.5}
  \caption{Best-fit parameters from the APO BG40 broadband and Gemini $r$-band and $i$-band datasets.
  Peak values for double-peaked distributions are reported together.
  Preferred solutions to the double-peaked parameters are bolded for clarity.
  We include the variance-weighted mean values across all filters for each parameter.}
	\label{tab:all_jktebop_table}
	\begin{tabular}{l c r r r}
    \hline
    \hline
		Parameter & BG40 broadband & $r$-band & $i$-band & Mean Value \\
		\hline
    \hline
    Light Ratio    &    0.020$^{+0.013}_{-0.010}$                      &  0.016$^{+0.006}_{-0.005}$    & 0.027$^{+0.010}_{-0.009}$  & {} \\
    $r_B + r_A$    &    0.118$^{+0.013}_{-0.010}$                      &  0.128$^{+0.008}_{-0.007}$    & 0.122$^{+0.011}_{-0.009}$  & {0.124$\pm$0.005} \\
    $r_B / r_A$    &    \textbf{0.449$\pm$0.015}                       &  \textbf{0.445$\pm$0.010}     & \textbf{0.439$\pm$0.014}   & \textbf{0.444$\pm$0.007} \\
    {}             &    0.573$\pm$0.027                                &  0.565$\pm$0.026              & 0.572$\pm$0.029            & {0.570$\pm$0.016} \\
    $i$ ($^\circ$) &    \textbf{87.9$\pm$0.4}                          &  \textbf{87.5$\pm$0.4}        & \textbf{87.7$\pm$0.5}      & \textbf{87.7$\pm$0.2} \\
    {}             &    86.4$\pm$0.4                                   &  86.0$\pm$0.3                 & 86.0$\pm$0.4               & {86.0$\pm$0.2} \\
    Period (d)     &    $0.0281258394\pm(1.5\times10^{-9})$            & {}                            & {}                         & {} \\
    \hline
    $R_A$ (${\rm R}_\odot$) &   0.029$\pm$0.010                        & {$0.032\pm0.009$}             & {$0.031\pm0.010$}          & {0.031$\pm$0.006} \\
    $R_B$ (${\rm R}_\odot$) &   \textbf{0.013$\pm$0.009}               & \textbf{0.014$\pm$0.008}      & \textbf{0.013$\pm$0.008}   & \textbf{0.013$\pm$0.005} \\
    {}                      &   0.017$\pm$0.009                        & {$0.018\pm0.009$}             & {$0.018\pm0.009$}          & {0.018$\pm$0.005}\\

    \hline
    \hline
	\end{tabular}
\end{table*}

%% file: gemini_r_corner_fig.tex
\begin{figure*}
\centering
\includegraphics[scale=0.25]{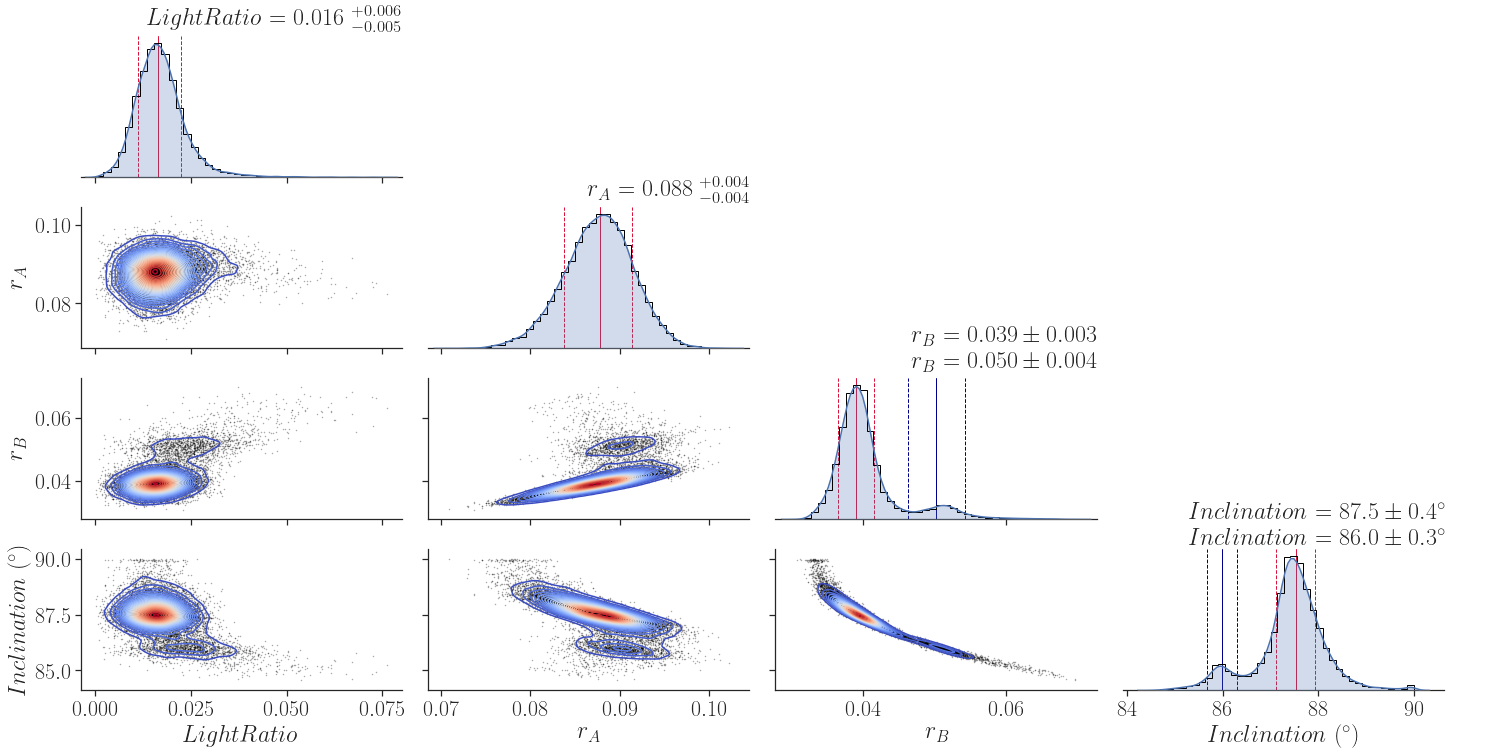}
\caption{Parameter distributions from Monte Carlo fits to the Gemini $r$-band light curve.
The diagonal contains the 1-D parameter distribution split into 40 bins (black histogram) with a 1-D Gaussian KDE overplotted as a blue shaded distribution.
The off-diagonal plots contain the 2-D parameter distributions with 2-D Gaussian KDE contours overplotted.
Primary/Secondary median fits and $1\sigma$ ranges are marked with vertical red/blue lines for single/double-peaked distributions.}
\label{fig:gemini_r_corner}
\end{figure*}

%% file: gemini_i_corner_fig.tex
\begin{figure*}
\centering
\includegraphics[scale=0.25]{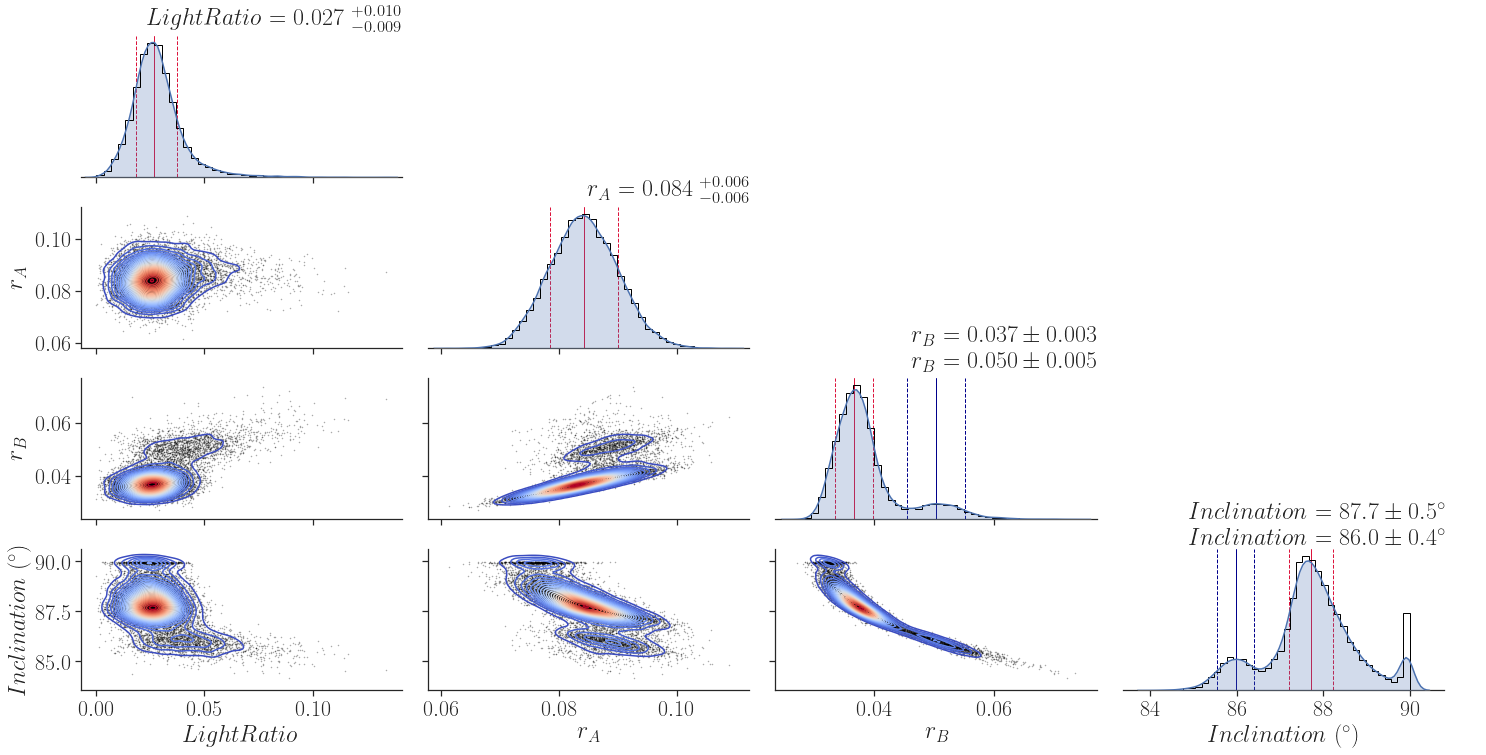}
\caption{Parameter distributions from Monte Carlo fits to the Gemini $i$-band light curve.
The diagonal contains the 1-D parameter distribution split into 40 bins (black histogram) with a 1-D Gaussian KDE overplotted as a blue shaded distribution.
The off-diagonal plots contain the 2-D parameter distributions with 2-D Gaussian KDE contours overplotted.
Primary/Secondary median fits and $1\sigma$ ranges are marked with vertical red/blue lines for single/double-peaked distributions.}
\label{fig:gemini_i_corner}
\end{figure*}

%% file: phased_model_lightcurves.tex
\begin{figure*}
\centering
\includegraphics[scale=0.3]{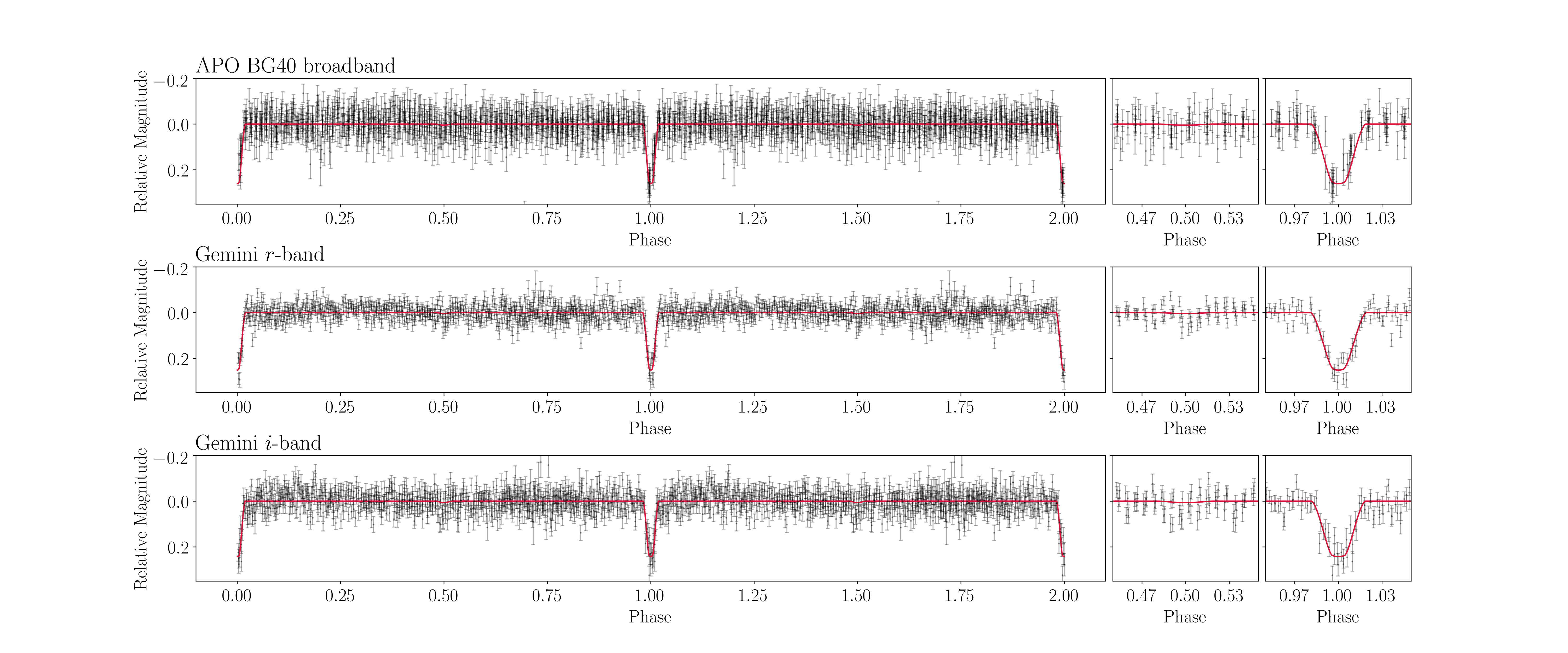}
\caption{Phase-folded light curves for APO BG40 broadband (top), Gemini $r$-band (middle), and Gemini $i$-band (bottom).
Best-fit models based on JKTEBOP Monte Carlo results are overplotted in red.
Zoomed-in plots surrounding the primary and secondary eclipses are included for each filter.
The secondary eclipse is not seen in any filter.}
\label{fig:phase_model_lightcurves}
\end{figure*}

%% file: eclipse_table.tex
\begin{table}
	\centering
  \renewcommand{\arraystretch}{1.5}
  \caption{Best-fit mid-eclipse times (barycentric dynamical time) for the APO BG40 broadband data.}
	\label{tab:eclipse}
	\begin{tabular}{c}
		\hline
		$T_0$ (BJD\_TDB) \\
		\hline
    {$2457814.82095\pm0.00005$} \\
    {$2458073.88809\pm0.00002$} \\
    {$2458132.89610\pm0.00003$} \\
    \hline
	\end{tabular}
\end{table}

%% file: apo_ephem_fig.tex
\begin{figure}
\centering
\includegraphics[scale=0.4]{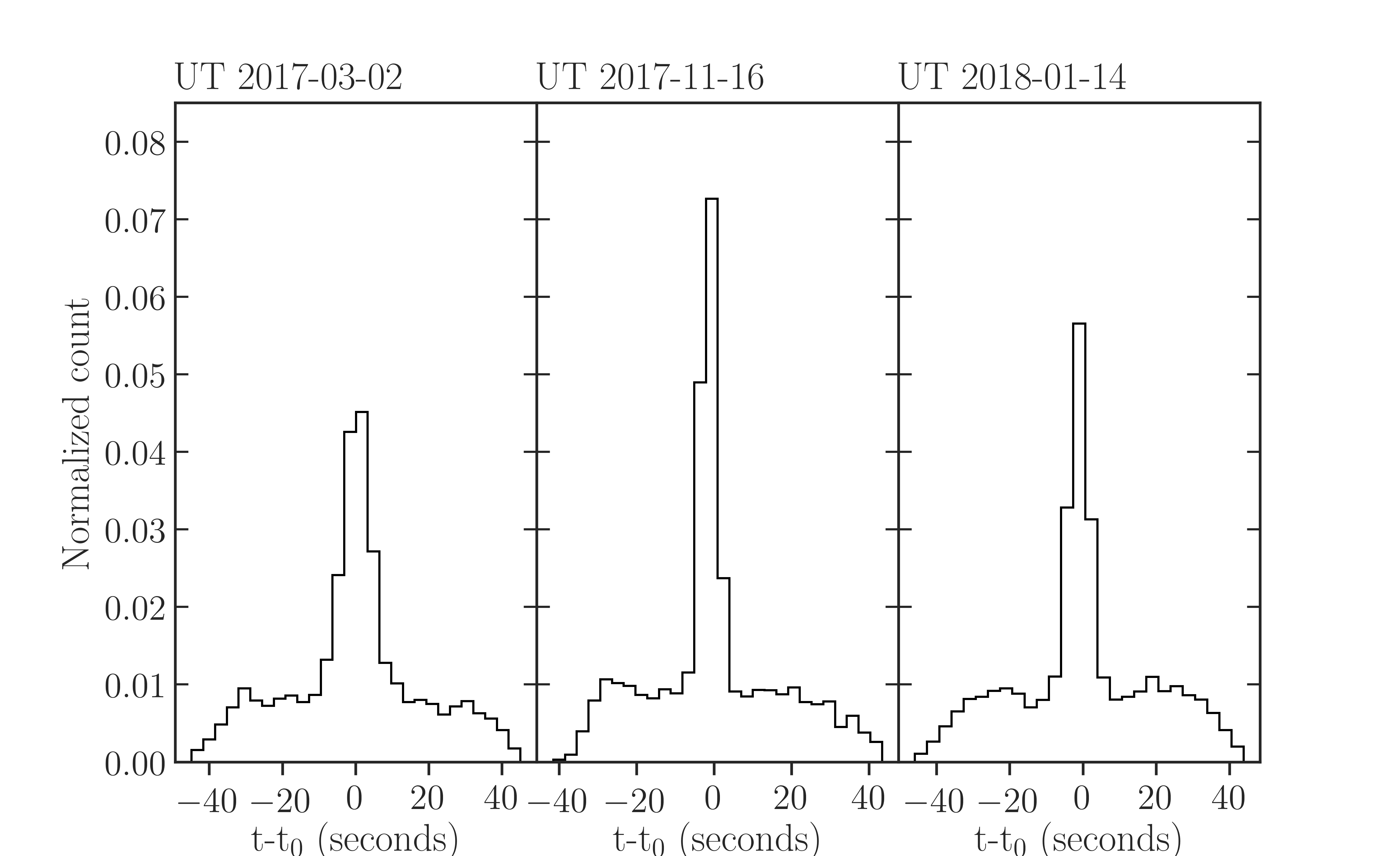}
\caption{Mid-eclipse timing distributions from 50,000 Monte Carlo fits to each of the three APO BG40 broadband light curves.
Fits that converged greater than 45 seconds from the median were excluded as unphysical.}
\label{fig:apo_ephem}
\end{figure}

%% file: grav_offset2.tex
\begin{figure}
\centering
\includegraphics[scale=0.4]{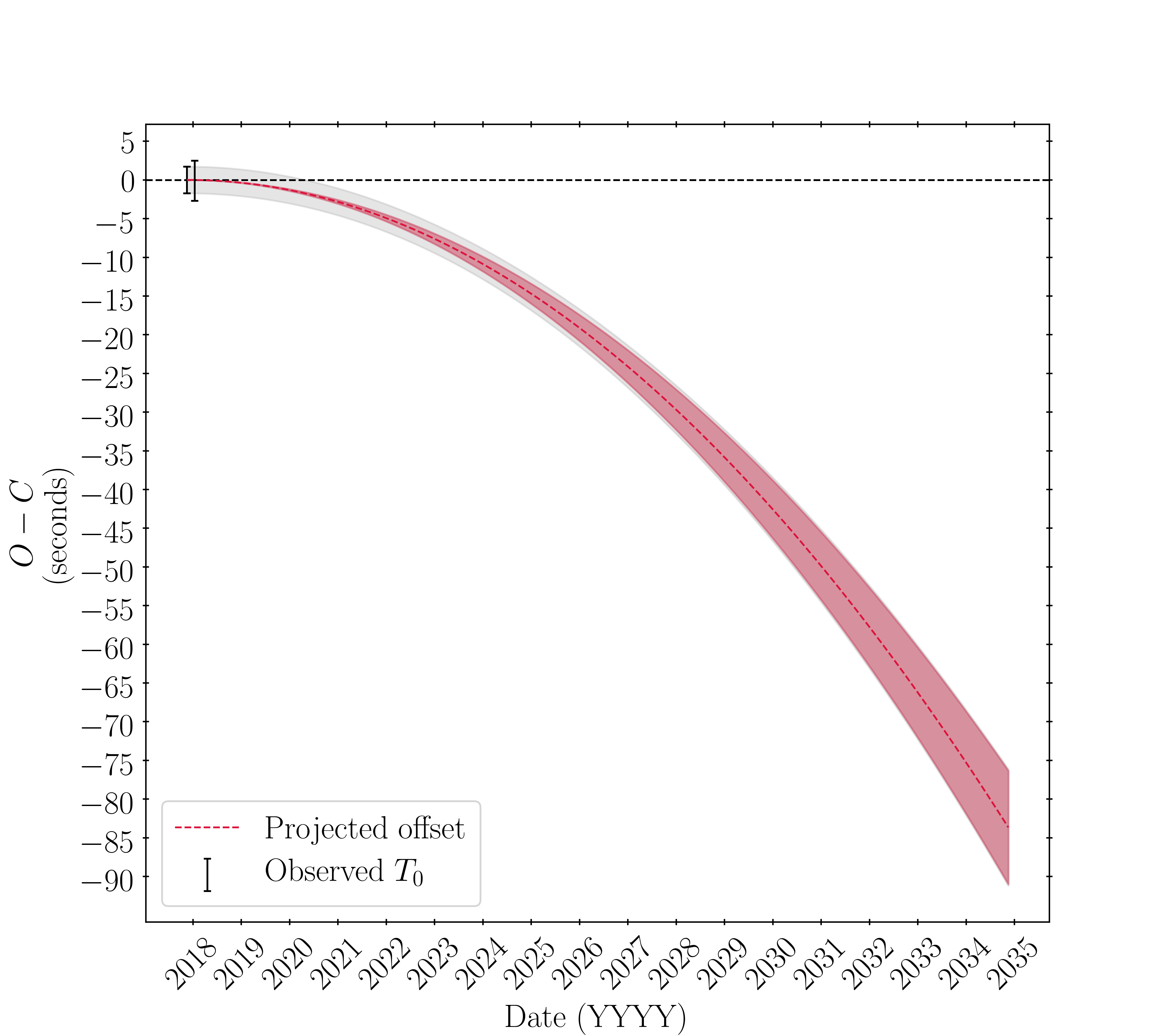}
\caption{Best-fit mid-eclipse timing measurements for J0822$+$3048 for our two new epochs of APO data discussed in the text
plotted as offsets to the expected mid-eclipse timing, in seconds, based on the period of $P=0.0281258394 {\rm d}$ determined through light curve fitting.
We include the projected offsets based on \citet{piro2011} estimates of angular momentum loss solely due to gravitational wave emission as
a function of time as a red dashed line with a shaded $1\sigma$ range up to the expected launch date of the LISA mission in 2034.
The dark grey shaded region represents the projected offset from gravitational wave emission, but also includes uncertainty in our initial time of eclipse measurement added in quadrature.
}
\label{fig:grav_offset}
\end{figure}

%% file: kosa2020b.bbl
\newpage
\begin{thebibliography}{}
\bibitem[Astropy Collaboration et al.(2013)]{astropy2013} Astropy Collaboration, Robitaille, T.~P., Tollerud, E.~J., et al.\ 2013, \aap, 558, A33
\bibitem[Astropy Collaboration et al.(2018)]{astropy2018} Astropy Collaboration, Price-Whelan, A.~M., Sip{\H{o}}cz, B.~M., et al.\ 2018, \aj, 156, 123
\bibitem[\protect\citeauthoryear{Benacquista}{2011}]{benacquista2011} Benacquista M.~J., 2011, ApJL, 740, L54
\bibitem[\protect\citeauthoryear{Brown et al.}{2011}]{brown2011} Brown W.~R., Kilic M., Hermes J.~J., Allende Prieto C., Kenyon S.~J., Winget D.~E., 2011, ApJL, 737, L23
\bibitem[\protect\citeauthoryear{Brown et al.}{2017}]{brown2017} Brown W.~R., Kilic M., Kosakowski A., Gianninas A., 2017, ApJ, 847, 10
\bibitem[\protect\citeauthoryear{Brown et al.}{2020}]{brown2020} Brown W.~R., Kilic M., Kosakowski A., Andrews J.~J., Heinke C.~O., Ag{\"u}eros M.~A., Camilo F., et al., 2020, ApJ, 889, 49
\bibitem[\protect\citeauthoryear{Burdge et al.}{2019a}]{burdge2019} Burdge K.~B., Coughlin M.~W., Fuller J., Kupfer T., Bellm E.~C., Bildsten L., Graham M.~J., et al., 2019a, Natur, 571, 528
\bibitem[\protect\citeauthoryear{Burdge et al.}{2019b}]{burdge2019b} Burdge K.~B., Fuller J., Phinney E.~S., van Roestel J., Claret A., Cukanovaite E., Gentile Fusillo N.~P., et al., 2019b, ApJL, 886, L12
\bibitem[\protect\citeauthoryear{Burdge et al.}{2020a}]{burdge2020a} Burdge K.~B., Prince T.~A., Fuller J., Kaplan D.~L., Marsh T.~R., Tremblay P.-E., Zhuang Z., et al., 2020a, arXiv, arXiv:2009.02567
\bibitem[\protect\citeauthoryear{Burdge et al.}{2020b}]{burdge2020b} Burdge K.~B., Coughlin M.~W., Fuller J., Kaplan D.~L., Kulkarni S.~R., Marsh T.~R., Prince T.~A., et al., 2020b, arXiv, arXiv:2010.03555
\bibitem[\protect\citeauthoryear{Claret et al.}{2020}]{claret2020} Claret A., Cukanovaite E., Burdge K., Tremblay P.-E., Parsons S., Marsh T.~R., 2020, A\&A, 634, A93
\bibitem[\protect\citeauthoryear{Coughlin et al.}{2020}]{coughlin2020} Coughlin M.~W., Burdge K., Phinney E.~S., van Roestel J., Bellm E.~C., Dekany R.~G., Delacroix A., et al., 2020, MNRAS, 494, L91
\bibitem[\protect\citeauthoryear{Drake et al.}{2010}]{drake2010} Drake A.~J., Beshore E., Catelan M., Djorgovski S.~G., Graham M.~J., Kleinman S.~J., Larson S., et al., 2010, arXiv, arXiv:1009.3048
\bibitem[\protect\citeauthoryear{Eastman, Siverd, \& Gaudi}{2010}]{eastman2010} Eastman J., Siverd R., Gaudi B.~S., 2010, PASP, 122, 935
\bibitem[\protect\citeauthoryear{Fontaine, Brassard, \& Bergeron}{2001}]{fontaine2001} Fontaine G., Brassard P., Bergeron P., 2001, PASP, 113, 409
\bibitem[\protect\citeauthoryear{Fuller \& Lai}{2013}]{fuller2013} Fuller J., Lai D., 2013, MNRAS, 430, 274. doi:10.1093/mnras/sts606
\bibitem[\protect\citeauthoryear{Gianninas et al.}{2013}]{gianninas2013} Gianninas A., Strickland B.~D., Kilic M., Bergeron P., 2013, ApJ, 766, 3
\bibitem[\protect\citeauthoryear{Hallakoun et al.}{2016}]{hallakoun2016} Hallakoun N., Maoz D., Kilic M., Mazeh T., Gianninas A., Agol E., Bell K.~J., et al., 2016, MNRAS, 458, 845
\bibitem[\protect\citeauthoryear{Hermes et al.}{2012}]{hermes2012} Hermes J.~J., Kilic M., Brown W.~R., Winget D.~E., Allende Prieto C., Gianninas A., Mukadam A.~S., et al., 2012, ApJL, 757, L21
\bibitem[\protect\citeauthoryear{Istrate et al.}{2016}]{istrate2016} Istrate A.~G., Marchant P., Tauris T.~M., Langer N., Stancliffe R.~J., Grassitelli L., 2016, A\&A, 595, A35
\bibitem[\protect\citeauthoryear{Kilic et al.}{2014}]{kilic2014} Kilic M., Hermes J.~J., Gianninas A., Brown W.~R., Heinke C.~O., Ag{\"u}eros M.~A., Chote P., et al., 2014, MNRAS, 438, L26
\bibitem[\protect\citeauthoryear{Kosakowski et al.}{2020}]{kosakowski2020} Kosakowski A., Kilic M., Brown W.~R., Gianninas A., 2020, ApJ, 894, 53
\bibitem[Landau \& Lifshitz(1958)]{landau1958} Landau, L.~D., \& Lifshitz, E.~M.,\ 1958, The Classical Theory of Fields, (Oxford: Oxford Pergamon Press)
\bibitem[\protect\citeauthoryear{Mukadam et al.}{2011}]{mukadam2011} Mukadam A.~S., Owen R., Mannery E., MacDonald N., Williams B., Stauffer F., Miller C., 2011, PASP, 123, 1423
\bibitem[\protect\citeauthoryear{Steinfadt et al.}{2010}]{steinfadt2010} Steinfadt J.~D.~R., Kaplan D.~L., Shporer A., Bildsten L., Howell S.~B., 2010, ApJL, 716, L146
\bibitem[\protect\citeauthoryear{Parsons et al.}{2011}]{parsons2011} Parsons S.~G., Marsh T.~R., G{\"a}nsicke B.~T., Drake A.~J., Koester D., 2011, ApJL, 735, L30
\bibitem[\protect\citeauthoryear{Parsons et al.}{2017}]{parsons2017} Parsons S.~G., G{\"a}nsicke B.~T., Marsh T.~R., Ashley R.~P., Bours M.~C.~P., Breedt E., Burleigh M.~R., et al., 2017, MNRAS, 470, 4473
\bibitem[\protect\citeauthoryear{Piro}{2011}]{piro2011} Piro A.~L., 2011, ApJL, 740, L53
\bibitem[\protect\citeauthoryear{Scott \& Howell}{2018}]{scott2018} Scott N.~J., Howell S.~B., 2018, SPIE, 10701, 107010G
\bibitem[\protect\citeauthoryear{Southworth, Maxted, \& Smalley}{2004}]{southworth2004} Southworth J., Maxted P.~F.~L., Smalley B., 2004, MNRAS, 351, 1277. doi:10.1111/j.1365-2966.2004.07871.x
\bibitem[\protect\citeauthoryear{Southworth et al.}{2005}]{southworth2005} Southworth J., Smalley B., Maxted P.~F.~L., Claret A., Etzel P.~B., 2005, MNRAS, 363, 529. doi:10.1111/j.1365-2966.2005.09462.x
\bibitem[\protect\citeauthoryear{Southworth}{2013}]{southworth2013} Southworth J., 2013, A\&A, 557, A119
\bibitem[\protect\citeauthoryear{Tremblay, Bergeron, \& Gianninas}{2011}]{tremblay2011} Tremblay P.-E., Bergeron P., Gianninas A., 2011, ApJ, 730, 128
\bibitem[\protect\citeauthoryear{Vennes et al.}{2011}]{vennes2011} Vennes S., Thorstensen J.~R., Kawka A., N{\'e}meth P., Skinner J.~N., Pigulski A., Ste\&{\textcommabelow s}acute, et al., 2011, ApJL, 737, L16

\end{thebibliography}
